\documentclass[12pt]{iopart}

\usepackage{textgreek}
\usepackage{algorithm, algorithmic}
\usepackage{verbatim}
\usepackage{graphicx}
\usepackage{dcolumn}
\usepackage{bm}
\usepackage{cite} 

\usepackage{xcolor}
\usepackage{ulem} 

\begin{document}
\title[Resource-efficient Bayesian protocol
for quantum magnetic field tracking]{Resource-efficient adaptive Bayesian tracking of \\ magnetic fields with a quantum sensor}

\author{K. Craigie, E. M. Gauger, Y. Altmann and C. Bonato}

\address{School of Engineering and Physical Sciences, SUPA, Heriot-Watt University, Edinburgh, EH14 4AS, UK}
\ead{klc31@hw.ac.uk, c.bonato@hw.ac.uk}

\begin{abstract}
Single-spin quantum sensors, for example based on nitrogen-vacancy centres in diamond, provide nanoscale mapping of magnetic fields. In applications where the magnetic field may be changing rapidly, total sensing time is crucial and must be minimised. Bayesian estimation and adaptive experiment optimisation can speed up the sensing process by reducing the number of measurements required. These protocols consist of computing and updating the probability distribution of the magnetic ﬁeld based on measurement outcomes and of determining optimized acquisition settings for the next measurement. However, the computational steps feeding into the measurement settings of the next iteration must be performed quickly enough to allow real-time updates. This article addresses the issue of computational speed by implementing an approximate Bayesian estimation technique, where probability distributions are approximated by a finite sum of Gaussian functions. Given that only three parameters are required to fully describe a Gaussian density, we find that in many cases, the magnetic field probability distribution can be described by fewer than ten parameters, achieving a reduction in computation time by factor 10 compared to existing approaches. For $T_2^* = 1~\mu s$, only a small decrease in computation time is achieved. However, in these regimes, the proposed Gaussian protocol outperforms the existing one in tracking accuracy.
\end{abstract}


\section{\label{Intro}Introduction}
Control and measurement of individual electron spins, achieved in the last two decades, enables highly sensitive measurements of magnetic fields \cite{Degen2017,Rondin2014,Levine2019}, with spatial resolution on the order of tens of nanometres \cite{Wrachtrup2016}. This is typically achieved through a point defect in diamond, the nitrogen-vacancy (NV) centre, which allows optical spin polarisation and readout even at room temperature \cite{Jelezko2006}. Capitalising on atomic-scale wavefunctions, the NV centre enables detection of static and periodic magnetic fields through the Zeeman shift on its electronic spin state with nanoscale resolution \cite{Balasubramanian2009,Bar-Gill2013}.

This capability has opened unprecedented opportunities in the field of condensed matter physics \cite{Casola2018}, such as the possibility to detect stray fields associated to currents in nanodevices \cite{Ku2020, Chang2017,Tetienne2017} and magnetisation in a variety of solid-state systems \cite{Thiel2016, Thiel2019, Yu2018, Rondin2013 ,Pelliccione2016}. NV centres can also be used as nanoscale sensors for temperature \cite{Kucsko2013}, strain \cite{Trusheim2016, Broadway2019} and electric ﬁelds \cite{Dolde2011,Michl2019}.  In addition to condensed matter systems, the biological compatibility of nanodiamonds containing NV centres allows monitoring of nanoscale, in-vivo processes\cite{Chipaux2018, McGuinness2011,Choi2020,Yukawa2020, Morita2020}.

Despite its success, one of the big challenges to further the deployment of NV sensing to practical applications is the data acquisition time. Room-temperature spin readout relies on the detection of a spin-dependent $\sim 30\%$ photoluminescence variation, on a signal which is typically (for a single NV) much less than one photon per readout shot. This results in the need of averaging over multiple repetitions, leading to long signal acquisition time \cite{Dinani2019}. A possible solution to this problem is to employ adaptive signal acquisition techniques, so that the experimental settings are optimised in real-time to minimise the detection time. In addition, sensing faster, or tracking a changing physical quantity in real-time could give new insight into previously inaccessible timescales.

Sensing of static magnetic fields with an NV centre can be carried out via an experiment made up of a series of measurements. It is the combination of many measurements that produces a reading for static magnetic field. The measurement parameters can be chosen optimally to narrow in on the magnetic field value sooner. This reduction in the number of measurements naturally allows for faster sensing.

Bayesian modeling and estimation, coupled with adaptive rules for experiment optimisation can be performed between measurements to select measurement parameters. Recent work has demonstrated the power of these techniques in speeding up the magnetic field sensing process \cite{Lumino2018, Zhang2018,Dinani2019, Cappellaro2012,VanDenBerg2020}. While several algorithms have been proposed and analysed \cite{Bonato2017,Santagati2019, Fiderer2020}, only one experimental implementation so far has enabled fully online operation \cite{Bonato2016}. An important point is that real-time implementation requires fast computations, on the microsecond scale, to optimise the settings for the following measurements. The computation timescales must be comparable to a single measurement time, otherwise it is more effective to simply keep taking measurements, without adaptive optimisation. For protocols relying on single-shot spin readout (at low temperatures), spin detection takes $\sim 10~ \mu$s, therefore the computation time should be shorter than this. Time-critical computation with minimal latency can be performed in parallel with a fast digital electronic system, such as a field-programmable gate array (FPGA). However, the number of (sequential) operations still needs to be kept as low as possible to keep the overall computational overhead real-time compatible.

Here we address this issue by adopting an approximate Bayesian estimation technique. In particular, at each point in time, we approximate the likelihood function and the posterior distribution of the parameter of interest as finite sums of Gaussian functions, i.e., Gaussian mixtures. Since Gaussian functions can be fully described by only three parameters (amplitude, centre and width), this allows faster processing as the number of parameters propagated over time is small compared to what would be required if the distributions were discretised on a grid or approximated via particle filtering \cite{Doucet2001}. We find that our Bayesian approach can typically be performed using only one or two Gaussian functions, achieving a 10-fold reduction in terms of computation time compared to previous non-approximate implementations\cite{Bonato2017,Santagati2019}.

While the work detailed here focuses on quantum sensing with NV centres in diamond, the protocol we examine can be readily applied to any other single-qubit quantum sensor \cite{Kraus2014,Yan2020}.

\section{\label{Background}Background}

A magnetic field applied to the electron spin induces a Zeeman splitting of the energy levels, which can be measured by a Ramsey experiment \cite{Degen2017}. In a Ramsey experiment, an equal spin superposition freely evolves under the applied magnetic field $B$, so that the spin eigenvalues acquire a relative phase, corresponding to a rotation at the Larmor frequency. The probability for outcome $\mu \in \{0;1\}$ given a Larmor frequency $f_B$ (corresponding to a magnetic field $B = f_B/\gamma$, where $\gamma$ is the gyromagnetic ratio, is
\newline\begin{equation}\label{eq:lik}
P(\mu|f_B,\theta,\tau) = \frac{1+e^{i\mu\pi}{e^{-(\frac{\tau}{T_2^*})^2}}cos(2\pi\tau f_B+\theta)}{2} \, .
\end{equation} where $\tau$ is the sensing time, $T_2^*$ is the coherence time of the NV centre and $\theta$ the rotation angle of the measurement basis (which is controlled by the control phase of the second $\pi/2$ pulse in the Ramsey measurement sequence). In equation \ref{eq:lik}, $P(x|y,z)$ denotes the distribution of $x$, conditioned on the value of $(y,z)$, ie, the distribution of $x$ given $y$ and $z$.

In this article, we consider the problem of tracking a magnetic field (through $f_B$) that fluctuates on  timescales longer than a single measurement time, but potentially shorter than a large number of repetitions of the measurement time. We assume the fluctuations to be described by a Wiener process yielding a sequence of Larmor frequency values that are generated as

\begin{equation}\label{eq:fb_generation}
f_B^{(t + \delta t)}= f_B^{(t)} + \kappa \textrm{d}W^{(t)} \, .
\end{equation}
The diffusion coefficient, $\kappa$, is a measure of the magnetic field rate of change and $\textrm{d}W^{(t)}$ is an inﬁnitesimal Wiener increment during a time inverval $\delta t$. This process is simulated by discretising the time axis to intervals of length $\tau_{min}$ (minimum sensing time), and generating a normal distribution with variance $\tau_{min}$. In this way, a changing Larmor frequency signal is generated that will act as the ground truth in tracking simulations.
More generally for any small time interval $\delta t$, equation \ref{eq:fb_generation}, leads to the following Gaussian random walk distribution 
\begin{equation}
\label{eq:random_walk}
  P\left(f_B^{(t + \delta t)}\bigg|f_B^{t},\kappa \right) \propto \exp\left[-\frac{\left(f_B^{(t + \delta t)}-f_B^{(t)}\right)^2}{2\delta t \kappa^2}\right]
\end{equation}
which will be used in the tracking algorithm described in the next section.

\subsection{Bayesian Estimation} 
Bayesian online tracking of $f_B$ involves achieving a real-time approximation of the probability distribution $f_B$. In turn this allow us to optimise the experimental settings through $\theta$ and $\tau$ in equation \ref{eq:lik}. After the $n$-th measurement, the (posterior) distribution $P(f_B^{(t_n)}|\boldsymbol{\mu}^{(t_n)},\theta_n,\tau_n)$, of $f_B^{(t_n)}$ is updated using Bayes' rule
\begin{equation}\label{eq:bayes} 
P(f_B^{(t_n)}|\boldsymbol{\mu}^{(t_n)},\theta_n,\tau_n) \propto P(\mu^{(t_n)}|f_B^{(t)},\theta_n,\tau_n)P(f_B^{(t_n)}|\boldsymbol{\mu}^{(t_{n-1})},\theta_n,\tau_n), 
\end{equation}
where $\boldsymbol{\mu}^{(t_n)}=\left\lbrace\mu^{(t_0)},\ldots,\mu^{(t_n)} \right\rbrace$. In equation \ref{eq:bayes}, $P(f_B^{(t_n)}|\boldsymbol{\mu}^{(t_{n-1})},\theta_n,\tau_n)$ acts as a prior distribution and can be obtained via 
\begin{eqnarray}\label{eq:prediction}
    P(f_B^{(t_n)}|\boldsymbol{\mu}^{(t_{n-1})},\theta_n,\tau_n)=\nonumber\\ \quad \quad \int  P\left(f_B^{(t)}\right|f_B^{(t_{n-1})},\kappa)P(f_B^{(t_{n-1})}|\boldsymbol{\mu}^{(t_{n-1})},\theta_n,\tau_n)\textrm{d}f_B^{(t_{n-1})},
\end{eqnarray} 
with $P(f_B^{(t_{n-1})}|\boldsymbol{\mu}^{(t_{n-1})},\theta_n,\tau_n)$ obtained via the Bayesian update after the $(n-1)$-th measurement. The update rules for $(\theta_n,\tau_n)$, are described in more detail in Sections \ref{phase} and \ref{sensingtime}.

\begin{figure*}
\centering
\includegraphics[scale=0.5]{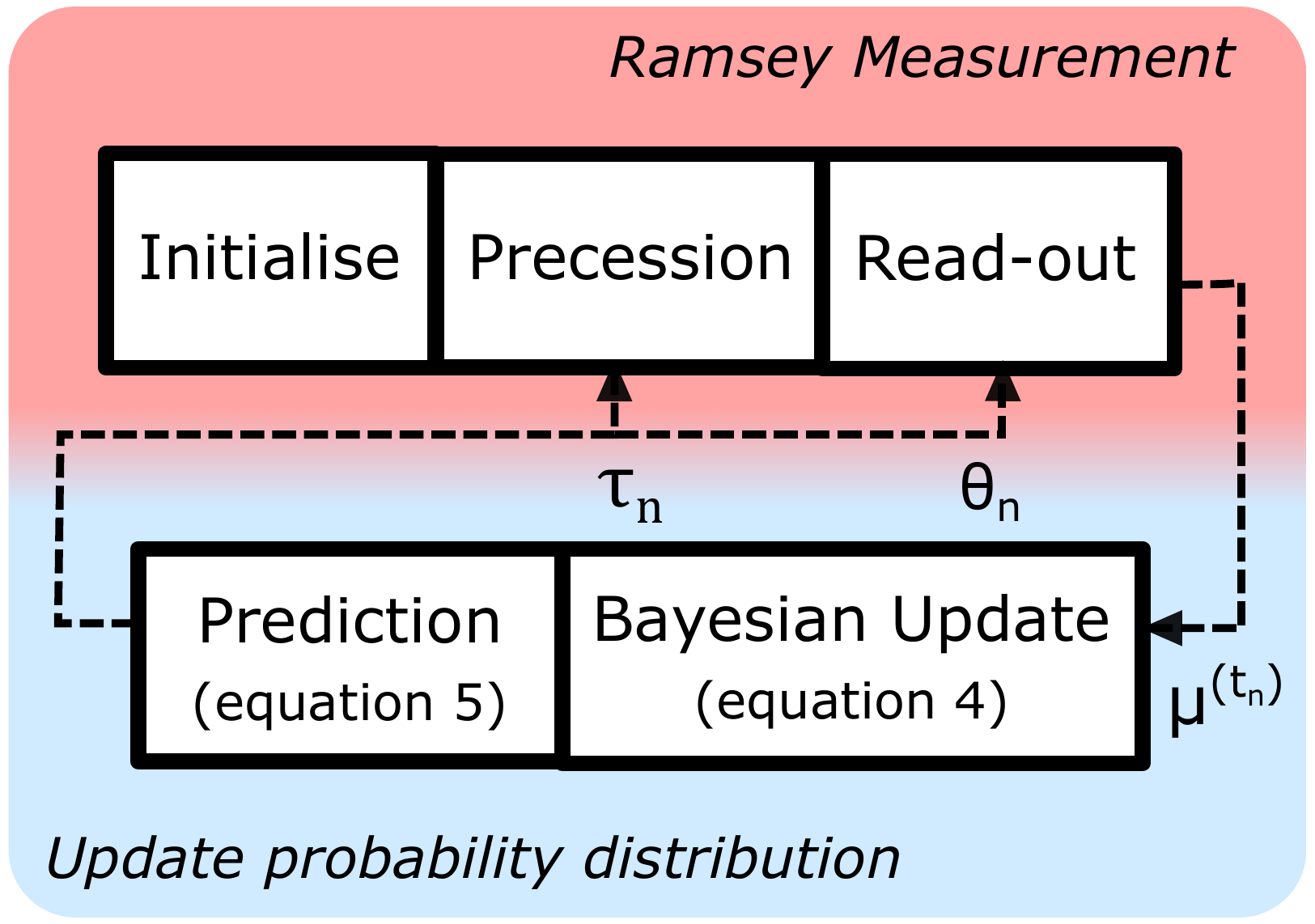}
\caption{\label{fig:overview}Graphical representation of Bayesian adaptive protocol, showing flow of information. The measurement outcome, $\mu^{(t_n)}$, from the Ramsey experiment is fed into the Bayesian update, followed by a prediction step. From this computation, the adaptive control phase, $\theta_n$, and sensing time, $\tau_n$, are determined. These values are fed into the next iteration of the Ramsey experiment.}
\end{figure*}

Figure \ref{fig:overview} illustrates the main principle of Bayesian adaptive tracking algorithms. However, as discussed in the introduction (section \ref{Intro}), exact Bayesian inference based on equations \ref{eq:bayes} and \ref{eq:prediction} is not tractable because of the shape of the likelihood in equation \ref{eq:lik}, which makes the integral in equation \ref{eq:prediction} not computationally amenable. While it is theoretically possible to use particle filters \cite{Doucet2001}, their computational complexity make them less attractive than approximate methods using Gaussian mixture approximations, as proposed here. 


\begin{figure*}
\centering
\includegraphics[scale=0.3]{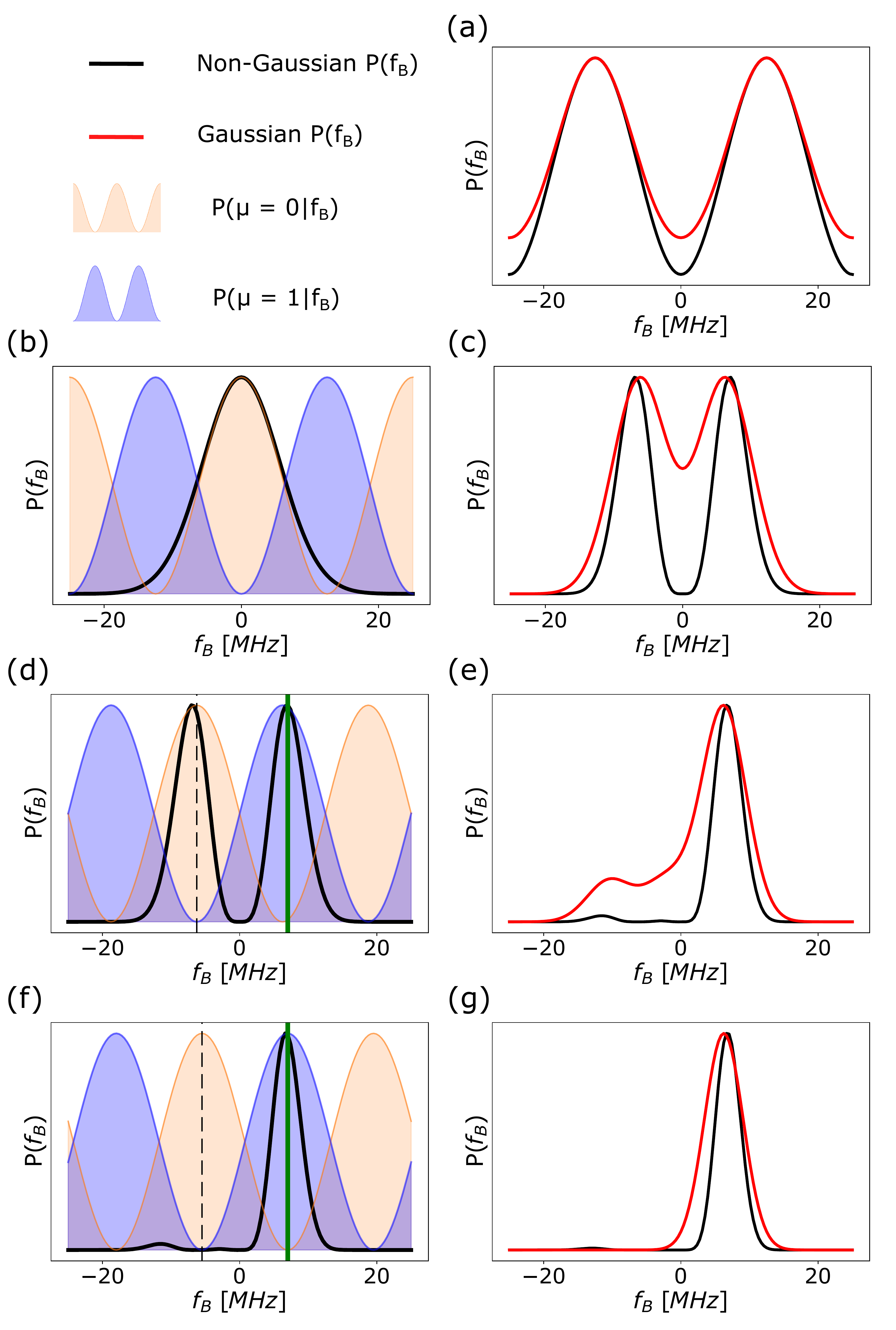}
\caption{\label{fig:trackingSteps} Step by step visualisation of Bayesian adaptive tracking protocol, showing the probability distribution for Larmor frequency updating after a change in magnetic field from 0 to 7 MHz. (a) likelihood function with Gaussian approximation, (b) prior and likelihood function for measurement n=1, (c) posterior for n=1, \textmu = 1, with Gaussian approximation, (d) prior and likelihood function for measurement n=2, (e) posterior for n=2, \textmu = 1, with Gaussian approximation, (f) prior and likelihood function for measurement n=1, (g) posterior for n=3, \textmu = 1, with Gaussian approximation. 
}
\end{figure*}

\section{\label{Method}Method}
In this article, we aim to reduce computational time by approximating $P(f_B^{(t_n)}|\boldsymbol{\mu}^{(t_n)},\theta_n,\tau_n)$ as a finite sum of Gaussian distributions. This is performed by approximating the likelihood $P(\mu^{(t_n)}|f_B^{(t_n)},\theta_n,\tau_n)$ by a weighted sum of Gaussian functions (with respect to $f_B^{(t_n)}$). Figure \ref{fig:trackingSteps}a shows how the actual cosine-shaped likelihood function is approximated by two shifted Gaussian peaks. As is apparent, the approximation is relatively accurate at the top of the peaks, but is rather poor at the bottom due to the Gaussian tails. In the following, we address the dual question of how well a magnetic field can be tracked despite this simplification, and examine the associated advantage of doing so in terms of the algorithm's reduction in computation time.

\subsection{Gaussian approximation of the likelihood function} 
We begin by approximating the initial likelihood in equation \ref{eq:lik} as:
\begin{equation} 
P(\mu|f_B,\theta_n,\tau_n) \approx \sum_{l=0}^{N_G} A e^{-\frac{(f_B-a_l)^2}{2\sigma_a^2}} \, .
\end{equation}

This approximation can be computed by a Taylor expansion of equation~\ref{eq:lik} at $2\pi\tau f_B+\theta_n+\mu\pi \approx 2 \pi l$ and re-writing as a Taylor expansion for an exponential. We chose to consider the case of no dephasing (its influence shall be discussed later in section \ref{dephasing}) and perfect readout fidelity as previous experiments \cite{Hensen2015} with NV centres at cryogenic temperatures have yielded a near-perfect readout fidelity of 0.96. Based on these two assumptions, the right-hand side of equation \ref{eq:lik} becomes:

\begin{eqnarray}
\frac{1+\cos(2\pi\tau_n f_B+\theta_n+\mu \pi)}{2} & \approx & 1-\frac{(2\pi\tau f_B+\theta_n+\mu \pi)^2}{4}\nonumber\\
& \approx & \exp^{-\left(\pi\tau_n f_B+\frac{\theta_n+\mu \pi}{2}\right)^2}. \end{eqnarray}

For the initial likelihood and under this approximation, we let all Gaussians have the same amplitude $A=1$ and width $\sigma_a = 1 / (\sqrt{2}\pi\tau)$, meaning they only differ in their centres $a_l$ as follows: 

\begin{equation}\label{eq.a} a_l = \frac{2\pi l+\pi \mu+\theta_n}{2\pi\tau} \, .
\end{equation}
This choice ensures that the displacement between adjacent Gaussians coincides with the period of the oscillatory likelihood function and that the Gaussian peaks align with the local maxima of the cosine function.

The number of periods of the likelihood function that fit in the prior frequency range is equal to the sensing time coefficient (the sensing time as a fraction of the minimum sensing time, $\tau_0$). This implies that the required number $N_G$ of Gaussians for the approximation is given by $2^N+1$, since this is the maximum sensing coefficient (plus one additional one to cater for edge peaks which are only partially visible). 

During tracking, only a handful of Gaussian peaks in the likelihood function overlap with the prior. Thus, the peaks that do not overlap with the prior do not need to included in the computation of the posterior distribution. We chose to cut out these unnecessary calculations by adapting the range of the likelihood function before each Bayesian update. Only peaks within $4(\sigma_a + \sigma_b)$ of any prior peak ($\sigma_b$ is the standard deviation of a Gaussian term in the prior distribution) were included in the computation of the posterior distribution.

\subsection{\label{Bayesian Update}Bayesian update}

Once the likelihood is approximated by a finite sum of Gaussians, the Bayesian update in equation~\ref{eq:bayes} becomes tractable and the posterior distribution reduced to a product of two mixtures of Gaussians (one arising from the prior distribution and one from the likelihood function). Defining ${A, a, \sigma_a}$ and ${B, b, \sigma_b}$ as, respectively, the amplitude, centre and standard deviation of two Gaussians, their product will be another Gaussian that is fully characterised by the parameters ${C, c, \sigma_c}$ with
\begin{eqnarray}\sigma_c & = \sqrt{\frac{\sigma_a^2\sigma_b^2}{\sigma_a^2+\sigma_b^2}} \, , \\
c & = \frac{a\sigma_b^2+b\sigma_a^2}{\sigma_a^2+\sigma_b^2} \, , \\
C &= A B e^{\frac{(a\sigma_b^2+b\sigma_a^2)^2/(\sigma_a^2+\sigma_b^2)-(a^2\sigma_b^2+b^2\sigma_a^2)}{(2\sigma_a^2\sigma_b^2)}} \, .
\end{eqnarray}
Note that after the initial assignment, the amplitudes of different Gaussians will no longer generally be identical, requiring calculation of $C$ above.

\subsection{Prediction Step}

As described in equation \ref{eq:prediction}, the (predictive) probability distribution of $f_B^{(t_n)}$ at time $t_n=t_{n-1} + \delta t_n$ can be found as the convolution of the probability distribution at time $t_{n-1}$ with a zero-mean Gaussian with variance $\delta t_n\kappa^2$. The time elapsed between measurements is given by $\delta t_n = \tau_n + t_{oh}$, where $\tau_n$ is the sensing time and $t_{oh}$ is the overhead time per measurement. This overhead time is added to account for the time it takes to physically carry out measurements in a lab. Naturally, this value varies depending on the precise experimental equipment and set-up. The distributions before and after prediction are therefore:
\begin{equation}
P(f_B^{(t_{n-1})}|\boldsymbol{\mu}^{(t_{n-1})},\theta_n,\tau_n) = \sum_{l} C_le^{-\frac{\left(f_B^{(t_{n-1})}-c_l\right)^2}{2\sigma^2_l}} \, ,
\end{equation}
\begin{equation}
P(f_B^{(t_n)}|\boldsymbol{\mu}^{(t_{n-1})},\theta_n,\tau_n) = \sum_{l} \frac{C_l\sigma_l}{\sqrt{\sigma^2_l + \kappa^2 dt}} e^{-\frac{\left(f_B^{(t_n)}-c_l\right)^2}{2(\sigma^2_l+\kappa^2 \delta t_n)}} \, ,
\end{equation}
where $\sigma'= \sqrt{\sigma^2 +\kappa^2\delta t}$ and $D = \frac{C\sigma}{\sigma'}$.

\subsection{Pruning the set of Gaussians}
It is important to note that, using mixtures of Gaussians to approximate the likelihood and the prior distribution (say with $m$ and $n$ terms, respectively), the number of Gaussians ($m \times n$) in the resulting posterior distribution keeps on increasing over time. To prevent this, we introduce an automatic pruning step based on amplitude thresholding. Any Gaussian with amplitude smaller than the predefined threshold is discarded.
\newline Additionally, we combine Gaussians that are very similar to each other, as defined by their Kullback-Leibler (KL) divergence \cite{Kullback1951}:
\newline
\begin{equation}KL(g_1,g_2) = \log\frac{\sigma_2}{\sigma_1} + \frac{\sigma_1^2+(a_1-a_2)^2}{2\sigma_2^2}-\frac{1}{2} \, ,
\end{equation}
where $g_1$ and $g_2$ are the two Gaussians being compared and $a_1,a_2$ are their respective central positions. 
When the KL divergence for a pair of Gaussians is below a given threshold, the pair is merged into one single Gaussian whose mean and variance are obtained by averaging the parameters of the two original Gaussians and whose amplitude is the sum of the original amplitudes. We found that the KL divergence threshold $KL_{th}=0.001$ for merging and the amplitude threshold $A_{th}=0.04$ for pruning work well in practice within the parameter range we have studied. 

The proposed tracking scheme might temporarily lose the track of $f_B$ and this yields all Gaussians having small amplitudes. If no amplitude exceeds the threshold, all the previous amplitudes and positions are kept unchanged and the variances are doubled. By simply broadening the previous distribution when the tracking is lost, the protocol picks up the signal again after a few iterations.

\subsection{Protocol Overview}

The proposed protocol is divided into phases, namely the initial adaptive sensing, followed by the adaptive tracking phase. The sensing portion is required to obtain a starting Larmor frequency value, which can then be tracked. Experimentally, this could also be obtained from locating the initial Larmor frequency by observing the spin resonance signal while sweeping the spin drive frequency. During sensing, the measurement time is not chosen adaptively, but in a predetermined sequence to narrow in on the Larmor frequency. Repeating each sensing time is required to minimise errors in sensing (see line 4 of algorithm \ref{Soverv}). However, the goal of adaptive sensing is to use the fewest Ramsey measurements and so a balance must be achieved. The number of Ramsey measurements performed at each sensing time is determined by integers G and F, which have been selected in accordance with previous work on adaptive sensing \cite{Bonato2016}. Due to this previous work studying in detail the initial sensing process, as well as the Gaussian approximation implementation providing an advantage only for tracking, this phase is not of primary interest for the current work. Algorithm \ref{Soverv} gives an overview of the initial sensing process, in which the posterior distribution of $f_B$ typically evolves from a uniform distribution (prior to any measurements) to an almost unimodal density which can be well approximated by a single Gaussian distribution. 

\begin{algorithm}
\caption{: Adaptive sensing overview.
Variables: adaptive control phase ($\theta_n$), measurement outcome ($\mu^{(t_n)}$), sensing time ($\tau_n$), minimum sensing time ($\tau_{min}$), number of sensing times (N). The remaining variables, $M_n$, G and F, define the number of repetitions for each sensing time.
}\label{Soverv}
\begin{algorithmic}[1]
\FOR{n= 0 to N-1}
\STATE $\tau_n = 2^n\tau_{min}$
\STATE $choose \ \theta_n$
\STATE  $M_n = G+F(n-1)$
\FOR{$ m=1\ to\ M_n:$}
\STATE $\mu^{(t_n)}= Ramsey(\theta= \theta_n, \tau=\tau_n)$
\STATE $Bayesian\_update(\mu=\mu^{(t_n)}, \theta = \theta_n^, \tau=\tau_n)$
\ENDFOR
\ENDFOR
\end{algorithmic}
\end{algorithm}
In contrast to the first phase, during the second phase of the protocol, the sensing time is chosen before each Ramsey measurement, based on the level of uncertainty of the current probability distribution of $f_B$. For completeness, a simplified example of the tracking protocol is provided in figure \ref{fig:trackingSteps}, where the adaptive control phase and sensing time are chosen to completely eliminate one of the two peaks in panel (d). Algorithm \ref{Toverv} illustrates the sequence of steps involved in adaptive tracking. In both tracking and sensing, the Gaussian approximation only affects steps ``choose $\theta_n$,'' ``Bayesian\_update'' and ``Prediction''. The other steps do not use $P(f_B^{(t)}|\boldsymbol{\mu}^{(t)},\theta_n,\tau_n)$ and thus remain unaffected. 
 
\begin{algorithm}
\caption{: Adaptive tracking overview.
Variables: adaptive control phase ($\theta_n$), measurement outcome ($\mu^{(t_n)}$), sensing time ($\tau_n$).
}\label{Toverv}
\begin{algorithmic}[1]
\FOR{total tracking time interval}
\STATE $choose \  \tau_n $
\STATE $choose \ \theta_n$
\STATE $\mu^{(t_n)}= Ramsey(\theta= \theta_n, \tau=\tau_n)$
\STATE $Bayesian\_update(\mu=\mu^{(t_n)}, \theta= \theta_n, \tau=\tau_n)$
\STATE $Prediction(\kappa)$
\ENDFOR
\end{algorithmic}
\end{algorithm}

\subsubsection{Adaptive Sensing Time}\label{sensingtime}

In this work, to determine whether the optimum sensing time has been chosen, we use the figure of merit proposed
in \cite{Bonato2017} and computed from an estimate of the standard deviation of the posterior distribution of $f_B$ (as detailed in equation 21 of \cite{Bonato2017}). This figure of merit is associated with a threshold above which the sensing time is judged insufficiently accurate and reduced by a factor two for the next measurement. This procedure is repeated until the threshold condition is met. If the figure of merit is lower than the threshold, the sensing time is instead increased by a factor two.

\subsubsection{Adaptive Control Phase}\label{phase}     

In order to maximise the information from each measurement, we adaptively set the angle of the measurement basis as \cite{Cappellaro2012}:

\begin{equation}\label{eq.5ph}
\theta_n=\frac{1}{2}arg\{p_{\,2t_n}\} \, ,
\end{equation}
where $p_{2t_n}$ is the prior probability distribution in Fourier space, $p_k$ for $k = 2t_n$, where $t_n$ is the sensing time coefficient. In terms of Gaussians, $p_{2t_n}$ is computed as:
\begin{equation}\label{eq.pk} 
p_{2t_n} = \sum_{l}^{}\sqrt{2\pi}C_l\sigma_l
e^{-2\pi^2(2t_n)^2\tau^2\sigma_l^2+l2\pi (2t_n)\tau c_l}.
\end{equation}
Other choices can be made for the optimal control phase, for example \cite{Lumino2018}: our Gaussian approximation will still work well.

\subsection{Considering dephasing}\label{dephasing}
The second exponential factor in the right-hand side of equation \ref{eq:lik} results from dephasing. It contributes a flat change in likelihood function amplitudes (which can be ignored due to normalisation), as well as introducing a positive $y$-shift ($P(\mu|f_B,\theta_n,\tau_n)$ increases and no longer dips to zero). Describing our probability distribution as a sum of Gaussians plus a single $y$-shift value would result in increased numbers of Gaussians during the Bayesian update. Without dephasing, a Bayesian update with $n_{lik}$ number of Gaussians in the likelihood function and $n_{pri}$ Gaussians in the prior would result in $n_{lik} n_{pri}$ Gaussians to describe the posterior. Taking into account dephasing and introducing a shift component in the same Bayesian update results in a posterior containing $n_{lik}n_{pri} + n_{lik} + n_{pri}$ Gaussians.

Though our pruning would remove many of these additional peaks, leading to a potential viability of this protocol, we have chosen not to include dephasing in our Gaussian adaptive protocol. As well as the shift being negligibly small for all our chosen sensing and $T_2^*$ times, the Gaussian approximation effectively introduces a much larger, unwanted y-shift in the same direction. This $y$-shift as a result of the Gaussian tails is the main cause of reduced tracking accuracy when implementing a Gaussian approximation.

\section{\label{Results}Results}
\begin{figure*}
\centering
\includegraphics[scale=0.8]{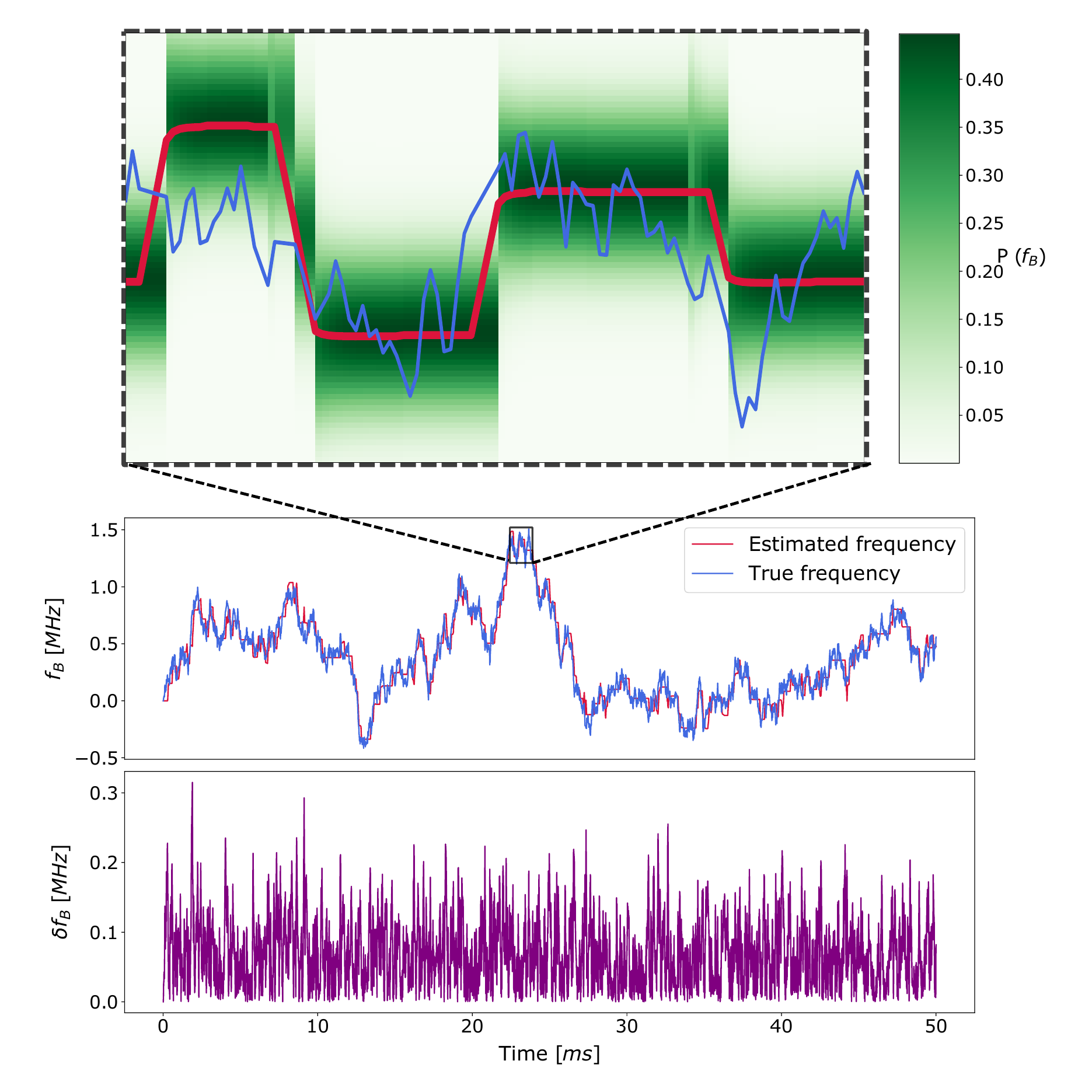}
\caption{\label{fig:results1} (a) example tracking of a changing Larmor frequency, performed by the Gaussian Bayesian adaptive protocol. Mean squared error of 0.087 MHz/ms, $\kappa = 10$ MHz H$z^{1/2}$ and $t_{oh} = 10~\mu s$ per measurement. Bottom subplot shows the difference between the estimation and ground truth. Green colour indicates the probability distribution at each point in time.}
\end{figure*}

To assess the performance of the Gaussian-approximation tracking protocol, we performed extensive numerical simulations using controlled settings, allowing for comparisons with ground truth parameters. We assume perfect spin readout fidelity and keep constant the amplitude pruning threshold $A_{th} = 0.04$, merging threshold $KL_{th}  = 0.001$ as well as $G = 5$ and $F=3$. The performance is assessed with the mean squared error (MSE)
\begin{equation}
\epsilon^2 = \frac{1}{T}\int_{0}^{T}|f_b-f_b^{est}|^2 \, ,
\end{equation}
where $f_{est}$ is the estimated frequency and $f_B$ the true frequency. Figure \ref{fig:results1} illustrates a successful tracking run, characterised by a typical MSE value of around 0.09 MHz/ms. Here, both the Gaussian and original method (i.e. with Gaussian approximation) produce the same quality of tracking (i.e., provide the same MSE) if the signal is successfully tracked. From visual examination of many runs, we decides to set the definition of a ``failed run" at an MSE of greater than 0.15 MHz/ms. This allows an empirical fail rate to be determined, which provides another useful measure of performance.

In these results, we performed two sorts of comparisons: a statistical and a direct comparison. The statistical comparison is based on running the tracking protocol many times, with each tracking run generating its own set of ``real" Larmor frequency values, i.e. the Gaussian and non-approximate method are not tested on identical signals. In our simulations, reusing a previously generated signal requires it to be significantly more discretised and so takes considerably more time to simulate. The time saved from using a different signal for each run allows us to sweep through a couple of experimental parameters in a manageable time frame. The overhead time, the extra experimental time taken to perform a Ramsey measurement, and $\kappa$, the prediction coefficient, were varied to examine the robustness of the protocol. During these runs, we set $T_2^*= 100~\mu{\rm s}$ and the interval over which the signal was tracked was varied so that each run would contain 1000 Ramsey measurements. 

The statistical comparisons give us a rough idea of speed and tracking performance through computational cost and MSE. To assess the improvement in terms of computational cost, we compared the average number of parameters used in the discretisation of the distribution of $f_B$. For the non-approximate implementation, this is the number of equally-space frequency points in the discretisation. For the Gaussian case, this is simply the number of Gaussian parameters i.e., three times the number of Gaussians employed for approximating the probability distribution.

In the direct comparisons, the Gaussian and non-approximate protocols are fed exactly the same signal to track multiple times. The interval over which the signal was tracked was set to 5~ms and the fail rate was used as the measure of tracking performance. The computation time was directly measured for the time-sensitive portion of the simulation (Bayesian update, prediction step and adaptive parameters, as these must be calculated in real time between Ramsey measurements). The computation time measurements were performed using Python's inbuilt \textit{time.perf\_counter} function, whilst running on a Lenovo ThinkPad E480 laptop, with Quad-core i7-8550U. The direct comparison results for a sample range of overhead times and $T_2^*$s can be found in table \ref{table}.

\begin{figure*}
\centering
\includegraphics[scale=0.85]{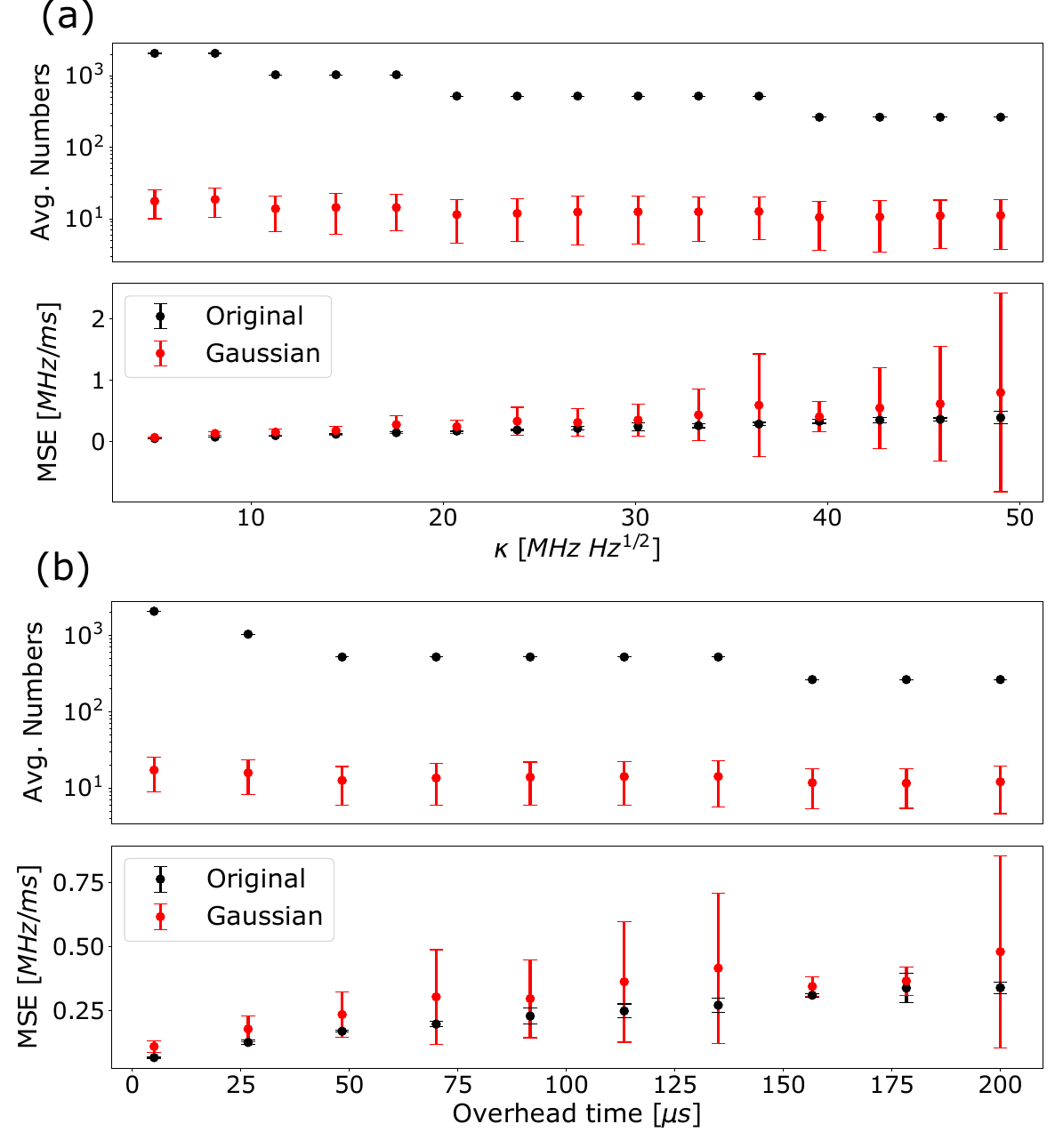}
\caption{\label{fig:results2} Both the original and Gaussian-approximate tracking protocols were tested for a range of overhead times and prediction coefficients (\textkappa). Each data point represents 1000 tracking runs, 5ms long. The bottom subplots are a measure of tracking performance (mean squared error) and the top subplots give the average numbers used to describe the Larmor frequency probability distribution (an indicator of relative computation speed) (a) sweeping \textkappa, overhead time fixed at $10\mu s$ per measurement. (b) sweeping overhead time, with $\kappa$ set to 10 MHz Hz$^{1/2}$.}
\end{figure*}

\begin{table}
\caption{\label{table}A direct comparison of 400, 5ms tracking runs on the same set of `true' Larmor frequency values for both Gaussian approximated and non-approximate algorithms. In each case, $\kappa = 10$~MHz Hz$^{1/2}$. F.R. is the fail rate, defined as the proportion of runs under an MSE of 0.15. The speed increase is the non-approximate protocol's computation time per measurement divided by the Gaussian computation time per measurement.}
\begin{indented}
\item[]\begin{tabular}{@{}lllll}
\br
$T_2^*$ ($\mu$s)&Overhead~($\mu$s)&Original F.R.~($\%$)&Gaussian F.R.~($\%$)& Speed increase\\
\mr
100&10&0.5&1&8.1\\
100&6&0&5.5&10.5\\
100&2&0.5&3&13.5\\
10&10&0.25&1.25&9.4\\
10&6&0.75&8.25&9.4\\
10&2&0.75&6.5&10.8\\
1&10&99.75&91.25&2.1\\
1&6&100&77.5&1.8\\
1&2&99.5&21&1.3\\
\br
\end{tabular}
\end{indented}
\end{table}

\subsection{Discussion}
As suggested by figure \ref{fig:trackingSteps}, without factoring computation time, the reliability of the tracking is generally degraded when using Gaussian tracking. In table \ref{table}, 5~ms of the exact same magnetic field fluctuations were tracked using both methods for  $\kappa = 10 MHz Hz^{1/2}$. This direct comparison was repeated 400 times to establish a fail rate. For the first entry in the table (overhead time $= 10\mu s$ and  $T_2^* = 100\mu s $), both methods produced very similar, near zero, fail rates. However, the fail rate for Gaussian runs is marginally higher than for those tracked using the original protocol. Table \ref{table} shows that this difference in fail rates is more pronounced for other values of overhead time and $T_2^*$. Interestingly, the Gaussian method performs worse when $T_2^*$ is reduced from 100\textmu s to 10\textmu s, where the original method sees almost no reduction in tracking quality.

On the other hand, the original method is incapable of tracking at $T_2^* = 1~\mu s$ with the chosen level of discretistation. However, the Gaussian protocol performs better, even being able to track the majority of runs when the overhead is $2~\mu s$. Upon further inspection of the MSE values, the improvements in tracking accuracy at $T_2^* = 1~\mu s$ are even more pronounced than indicated by the fail rates given in table \ref{table}. The Gaussian protocol's MSE values are comparatively much lower. To illustrate this, if we were to increase the failed run threshold by only 0.5MHz/ms, the fail rates for $10\mu s,6\mu s$,and $2\mu s$ would be 20\%, 10\% and 3.5\% respectively. Meanwhile, the fail rates for the original method would stay in the 90\%s. 

It is possible to improve the original method's tracking performance at $T_2^* = 1~\mu s$ by discretising the probability distribution further, though this naturally increases computation time. However, in our test simulations, the original method never matched the Gaussian in tracking performance, even when discretising the original method to the point that it was 200 times slower than the Gaussian method. 

For the statistical comparisons, we chose to vary overhead time and prediction coefficient, $\kappa$, as they had been used to benchmark previous comparisons of adaptive versus non-adaptive schemes \cite{Bonato2017}. In this way, we could test the robustness of the methods, as these are also not parameters we have control over in an experiment. From subplots \ref{fig:results2}b and c, it can be seen that the mean squared error tends to increase as these variables increase. We also see that the Gaussian method is overall more likely to break down at larger values of $\kappa$ and overhead time, than the smaller values. This is because, at larger values of both these variables, we can fit in fewer measurements per change of magnetic field. We have less time to narrow in on the correct frequency, and this puts the Gaussian method at a disadvantage, since we cannot factor in the potential computational time reduction in the simulation.

The average numbers is a reflection of the Bayesian update computation time. In the case of the original method, this is mostly in the thousands. Note that these numbers are not constant as the method discretises of the Larmor frequency probability distribution with a grid whose resolution is proportional to the number of sensing times $N$. Moreover, $N$ varies depending on $\kappa$ and $t_{oh}$ since the maximum sensing time is optimised according to equation 14 of \cite{Bonato2017}. For the Gaussian method though, the average number of parameters is around eight or nine for this length of tracking run, suggesting a substantial reduction in computational complexity and thus a gain in computational time. The average number of parameters is dragged up by the initial sensing, which initially requires hundreds of Gaussian parameters to describe it. The tracking uses mostly three, sometimes six, and occasionally nine or more parameters, indicating that a handful of Gaussian peaks are largely sufficient and deliver adequate performance. As we have shown, this translates to a reduction in the number of operations required to track a changing magnetic field. This is confirmed by the direct comparison time tests, in which the Gaussian tracking was found to be around an order of magnitude faster for the majority of our sample tests in table \ref{table}. 

The outliers in tracking speed improvements are again at $T_2^* = 1\mu s$, with only slight improvement occurring. However, as discussed above, the Gaussian method performs tracking better than the non-approximate method in this regime. Here, the improved performance, in combination with the improvement in speed makes choosing the Gaussian protocol worthwhile. As an aside, this smaller speed increase is partly due to the Gaussian protocol slowing down. When the Gaussian protocol is uncertain of the Larmor frequency, it will naturally use more peaks to describe the distribution, increasing computation time. In this way, the Gaussian protocol increases its tracking performance by sacrificing computation speed.

\section{Conclusion}
In this article, we have simulated and tracked a fluctuating magnetic field via Ramsey experiments on an NV centre. Ramsey measurements were optimised through an adaptive Bayesian update protocol, with Gaussian approximation of all probability distributions. A comparison of the Gaussian-approximated with the original protocol revealed that the approximation yielded potentially significant increases in computation speed, which ranged from 1.3 to 13.5 times faster, depending on the experiment parameters. In the cases of $T_2^* = 100~\mu s, 10~\mu s$, the approximation performed slightly poorer at tracking than the non-approximated protocol. However, these cases yielded larger increases in speed -- around an order of magnitude faster than the original protocol. In the slower cases (where  $T_2^* = 1~\mu s$), this lesser increase in speed was compensated for by the Gaussian method having the better tracking performance. This protocol could find applications in sensing settings where one needs to track a fluctuating signal with statistics that are, at least approximately, known in advance. 

For example, one could use an NV centre in a nanodiamond to monitor temperature drift inside a living cell \cite{Choi2020,Yukawa2020}. Measuring temperature is an integral part of studying energy metabolism \cite{Meyer2017} or developmental processes \cite{Begasse2015, Chong2018}.  One other issue with nanodiamonds, is that while moving in a fluid medium, they rotate considerably, often very rapidly. Our protocol could be extended to track this rotation with minimal resource consumption.

Another possible application is in experiments with levitated nanodiamond, where our technique could be used for fast tracking of rotation. In these experiments, the nanodiamond containing NV centres is held in place translationally using ion traps \cite{Quidant2018,Kuhlicke2014}, optical traps \cite{Frangeskou2018} or magnetic traps \cite{Hsu2016}. The librational (rotational) frequencies of trapped nanodiamond vary from 100s \cite{Delord2020} of Hz to 1GHz \cite{Reimann2018}. For the lower frequencies, in which tracking resolutions of 1ms per data point are suitable, our method could be directly applied. However, the method could also be used in conjunction with ac magnetomety \cite{Maze2008,DeLange2011}, using spin-echo instead of Ramsey measurements, to achieve tracking of faster, periodic librations.

Tracking provides the information required for realigning the nanodiamond orientation, for whichever feedback mechanism the traps use. This feedback mechanism could potentially be 3D Helmholtz coils in the case of ion traps or optical traps. For a magnetic trap, it has been proposed \cite{Plenio2019} that the diamond orientation be confined with an electrode using the dielectric force on the non-spherical diamond. Though this method does not conventionally require feedback, tracking may nonetheless prove useful in testing the confinement. 

Finally, the Gaussian-approximation described here could be applied to track a quantum signal, such as the magnetic field arising from a bath of nuclear spins surrounding a central electron spin. Previous theoretical work has shown that, by adaptively tracking the fluctuating nuclear magnetic field and narrowing its distribution through the back-action of the quantum measurement process, one can considerably extended the coherence time of the central spin \cite{Scerri2020}. The protocol described here can reduce the computational complexity of this task, enabling faster and more precise tracking.

\ack
The authors would like to thank Eleanor Scerri for helpful discussions. This project is supported by the Engineering and Physical Sciences Research council through grants EP/S000550/1 and EP/T01377X/1, and by a Weizmann-UK Joint Research Program grant.  K.C. acknowledges studentship funding from EPSRC under grant no. EP/L015110/1. Y.A. is supported by the Royal Academy of Engineering under the Research Fellowship scheme RF201617/16/31.

\section*{References}
\bibliographystyle{iopart-num}
\bibliography{references}

\providecommand{\noopsort}[1]{}\providecommand{\singleletter}[1]{#1}%
\providecommand{\newblock}{}
\begin{thebibliography}{10}
\expandafter\ifx\csname url\endcsname\relax
  \def\url#1{{\tt #1}}\fi
\expandafter\ifx\csname urlprefix\endcsname\relax\def\urlprefix{URL }\fi
\providecommand{\eprint}[2][]{\url{#2}}

\bibitem{Degen2017}
Degen C~L, Reinhard F and Cappellaro P 2017 {\em Rev. Mod. Phys.\/} {\bf 89}
  035002

\bibitem{Rondin2014}
Rondin L, Tetienne J~P, Hingant T, Roch J~F, Maletinsky P and Jacques V 2014
  {\em Rep. Prog. Phys.\/} {\bf 77} 056503

\bibitem{Levine2019}
Levine E~V, Turner M~J, Kehayias P, Hart C~A, Langellier N, Trubko R, Glenn
  D~R, Fu R~R and Walswortho R~L 2019 {\em Nanophotonics\/} {\bf 8} 1945

\bibitem{Wrachtrup2016}
Wrachtrup J and Finkler A 2016 {\em J. Magn. Reson.\/} {\bf 269} 225

\bibitem{Jelezko2006}
Jelezko F and Wrachtrup J 2006 {\em physica status solidi (a)\/} {\bf 203}
  3207--3225 ISSN 18626300

\bibitem{Balasubramanian2009}
Balasubramanian G, Neumann P, Twitchen D, Markham M, Kolesov R, Mizuochi N,
  Isoya J, Achard J, Beck J, Tissler J, Jacques V, Hemmer P~R, Jelezko F and
  Wrachtrup J 2009 {\em Nature Materials\/} {\bf 8} 383--387 ISSN 14764660

\bibitem{Bar-Gill2013}
Bar-Gill N, Pham L~M, Jarmola A, Budker D and Walsworth R~L 2013 {\em Nature
  Communications\/} {\bf 4} 1--6 ISSN 20411723

\bibitem{Casola2018}
Casola F, van~der Sar T and Yacoby A 2018 {\em Nature Reviews Materials\/} {\bf
  3} 17088 (\textit{Preprint} \eprint{1804.08742})

\bibitem{Ku2020}
Ku M~J, Zhou T~X, Li Q, Shin Y~J, Shi J~K, Burch C, Anderson L~E, Pierce A~T,
  Xie Y, Hamo A, Vool U, Zhang H, Casola F, Taniguchi T, Watanabe K, Fogler
  M~M, Kim P, Yacoby A and Walsworth R~L 2020 {\em Nature\/} {\bf 583} 537--541
  ISSN 14764687 (\textit{Preprint} \eprint{1905.10791})

\bibitem{Chang2017}
Chang K, Eichler A, Rhensius J, Lorenzelli L and Degen C~L 2017 {\em Nano
  Letters\/} {\bf 17} 2367--2373 ISSN 15306992 (\textit{Preprint}
  \eprint{1609.09644})

\bibitem{Tetienne2017}
Tetienne J~P, Dontschuk N, Broadway D~A, Stacey A, Simpson D~A and Hollenberg
  L~C 2017 {\em Science Advances\/} {\bf 3} e1602429 ISSN 23752548

\bibitem{Thiel2016}
Thiel L, Rohner D, Ganzhorn M, Appel P, Neu E, Kleiner R, Koelle D and
  Maletinsky P 2016 {\em Nature Nanotechnology|\/} {\bf 11}

\bibitem{Thiel2019}
Thiel L, Wang Z, Tschudin M~A, Rohner D, Guti{\'{e}}rrez-Lezama I, Ubrig N,
  Gibertini M, Giannini E, Morpurgo A~F and Maletinsky P 2019 {\em Science\/}
  {\bf 364} 973--976 (\textit{Preprint} \eprint{1807.04898})

\bibitem{Yu2018}
Yu G, Jenkins A, Ma X, Razavi S~A, He C, Yin G, Shao Q, l~He Q, Wu H, Li W,
  Jiang W, Han X, Li X, Jayich A~C~B, Amiri P~K and Wang K~L 2018 {\em Nano
  Letters\/} {\bf 18} 980--986 ISSN 15306992

\bibitem{Rondin2013}
Rondin L, Tetienne J~P, Rohart S, Thiaville A, Hingant T, Spinicelli P, Roch
  J~F and Jacques V 2013 {\em Nature Communications\/}

\bibitem{Pelliccione2016}
Pelliccione M, Jenkins A, Ovartchaiyapong P, Reetz C, Emmanouilidou E, Ni N and
  {Bleszynski Jayich} A~C 2016 {\em Nature Nanotechnology\/} {\bf 11} 700--705
  ISSN 17483395 (\textit{Preprint} \eprint{1510.02780})

\bibitem{Kucsko2013}
Kucsko G, Maurer P~C, Yao N~Y, Kubo M, Noh H~J, Lo P~K, Park H and Lukin M~D
  2013 {\em Nature\/} {\bf 500} 54--58

\bibitem{Trusheim2016}
Trusheim M~E and Englund D 2016 {\em New J.Phys.\/} {\bf 18} 123023

\bibitem{Broadway2019}
Broadway D~A, Johnson B~C, Barson M~S~J, Lillie S~E, Dontschuk N, McCloskey
  D~J, Tsai A, Teraji T, Simpson D~A, Stacey A, McCallum J~C, Bradby J~E,
  Doherty M~W, Hollenberg L~C~L and Tetienne J~P 2019 {\em Nano Lett.\/} {\bf
  19, 7} 4543--4550

\bibitem{Dolde2011}
Dolde F, Fedder H, Doherty M~W, Nöbauer T, Rempp F, Balasubramanian G, Wolf T,
  Reinhard F, Hollenberg L~C~L, Jelezko F and Wrachtrup J 2011 {\em Nature
  Physics\/} {\bf 7} 459–463

\bibitem{Michl2019}
Michl J, Steiner J, Denisenko A, Bulau A, Zimmermann A, Nakamura K, Sumiya H,
  Onoda S, Neumann P, Isoya J and Wrachtrup J 2019 {\em Nano Lett. 2019\/} {\bf
  19} 4904--4910

\bibitem{Chipaux2018}
Chipaux M, van~der Laan K~J, Hemelaar S~R, Hasani M, Zheng T and Schirhagl R
  2018 {\em Small\/} {\bf 14} 1704263

\bibitem{McGuinness2011}
McGuinness L~P, Yan Y, Stacey A, Simpson D~A, Hall L~T, Maclaurin D, Prawer S,
  Mulvaney P, Wrachtrup J, Caruso F, Scholten R~E and Hollenberg L~C~L 2011
  {\em Nature Nanotech\/} {\bf 6} 358–363

\bibitem{Choi2020}
Choi J, Zhou H, Landig R, Wu H~Y, Yu X, Kucsko G, Maurer P, Needleman D, Samuel
  A~D~T, Park H and Lukin M~D 2020 {\em Proceedings of the National Academy of
  Sciences\/} {\bf 117,26} 201922730

\bibitem{Yukawa2020}
Yukawa H, Fujiwara M, Kobayashi K, Kumon Y, Miyaji K, Nishimura Y, Oshimi K,
  Umehara Y, Teki Y, Iwasaki T, Hatano M, Hashimoto H and Baba Y 2020 {\em
  Nanoscale Adv.\/} {\bf 2} 1859--1868

\bibitem{Morita2020}
Morita A, Nusantara A~C, Martinez F~P~P, Hamoh T, Damle V~G, van~der Laan K~J,
  Sigaeva A, Vedelaar T, Chang M, Chipaux M and Schirhagl R 2020 {Quantum
  monitoring the metabolism of individual yeast mutant strain cells when aged,
  stressed or treated with antioxidant} (\textit{Preprint} \eprint{2007.16130})

\bibitem{Dinani2019}
Dinani H~T, Berry D~W, Gonzalez R, Maze J~R and Bonato C 2019 {\em Phys. Rev.
  B\/} {\bf 99} 125413

\bibitem{Lumino2018}
Lumino A, Polino E, Rab A~S, Milani G, Spagnolo N, Wiebe N and Sciarrino F 2018
  {\em Phys Rev. Applied\/} {\bf 10} 044033

\bibitem{Zhang2018}
Zhang Y~H and Yang W 2018 {\em New J. Phys.\/} {\bf 20} 093011

\bibitem{Cappellaro2012}
Cappellaro P 2012 {\em Phys. Rev. A\/} {\bf 85} 030301(R)

\bibitem{VanDenBerg2020}
{Van Den Berg} E 2020 {Efficient Bayesian phase estimation using mixed priors}
  (\textit{Preprint} \eprint{2007.11629v1})

\bibitem{Bonato2017}
Bonato C and Berry D~W 2017 {\em Phys Rev A\/} {\bf 95} 052348

\bibitem{Santagati2019}
Santagati R, Gentile A~A, Knauer S, Schmitt S, Paesani S, Granade C, Wiebe N,
  Osterkamp C, McGuinness L~P, Wang J, Thompson M~G, Rarity J~G, Jelezko F and
  Laing A 2019 {\em Phys. Rev. X\/} {\bf 9} 021019

\bibitem{Fiderer2020}
Fiderer L~J, Schuff J and Braun D 2020 {Neural-Network Heuristics for Adaptive
  Bayesian Quantum Estimation} (\textit{Preprint} \eprint{2003.02183v1})

\bibitem{Bonato2016}
Bonato C, Blok M~S, Dinani H~T, Berry D~W, Markham M~L, Twitchen D~J and Hanson
  R 2016 {\em Nature Nanotech\/} {\bf 11} 247–252

\bibitem{Doucet2001}
Doucet A, Smith A, {de Freitas} N and Gordon N 2001 {\em Sequential Monte Carlo
  Methods in Practice\/} Information Science and Statistics (Springer New York)
  ISBN 9780387951461

\bibitem{Kraus2014}
Kraus H, Soltamov V~A, Fuchs F, Simin D, Sperlich A, Baranov P~G, Astakhov G~V
  and Dyakonov V 2014 {\em Scientific Reports\/} {\bf 4} ISSN 20452322

\bibitem{Yan2020}
Yan F~F, Yi A~L, Wang J~F, Li Q, Yu P, Zhang J~X, Gali A, Wang Y, Xu J~S, Ou X,
  Li C~F and Guo G~C 2020 {\em npj Quantum Information\/} {\bf 6} 1--6 ISSN
  20566387

\bibitem{Hensen2015}
Hensen B, Bernien H, Drea{\'{u}} A~E, Reiserer A, Kalb N, Blok M~S, Ruitenberg
  J, Vermeulen R~F, Schouten R~N, Abell{\'{a}}n C, Amaya W, Pruneri V, Mitchell
  M~W, Markham M, Twitchen D~J, Elkouss D, Wehner S, Taminiau T~H and Hanson R
  2015 {\em Nature\/} {\bf 526} 682--686 ISSN 14764687

\bibitem{Kullback1951}
Kullback S and Leibler R~A 1951 {\em Ann. Math. Statist.\/} {\bf 22,1} 79--86

\bibitem{Meyer2017}
Meyer C~W, Ootsuka Y and Romanovsky A~A 2017 {\em Frontiers in Physiology\/}
  {\bf 8} 520

\bibitem{Begasse2015}
Begasse M~L, Leaver M, Vazquez F, Grill S~W and Hyman A~A 2015 {\em Cell
  Rep.\/} {\bf 10} 647

\bibitem{Chong2018}
Chong J, Amourda C and Saunders T~E 2018 {\em J. R. Soc. Interface.\/} {\bf 15}
  20180304

\bibitem{Quidant2018}
Conangla G~P, Schell A~W, Rica R~A and Quidant R 2018 {\em Nano Lett.\/} {\bf
  18} 3956--3961

\bibitem{Kuhlicke2014}
Kuhlicke A, Schell A~W, Zoll J and Benson O 2014 {\em Appl.Phys.Lett.\/} {\bf
  105} 073101

\bibitem{Frangeskou2018}
Frangeskou A~C, Rahman A~T~M~A, Gines L, Manda S, Williams O~A, Barker P~F and
  Morley G~W 2018 {\em New J.Phys.\/} {\bf 20} 043016

\bibitem{Hsu2016}
Hsu J~F, Ji P, Lewandowski C~W and D’Urso B 2016 {\em Sci. Rep.\/} {\bf 6}
  30125

\bibitem{Delord2020}
Delord T, Huillery P, Nicolas L and Hétet G 2020 {\em Nature\/} {\bf 580}
  56–59

\bibitem{Reimann2018}
Reimann R, Doderer M, Hebestreit E, Diehl R, Frimmer M, Windey D, Tebbenjohanns
  F and Novotny L 2018 {\em Physical Review Letters\/} {\bf 121} 033602 ISSN
  10797114

\bibitem{Maze2008}
Maze J~R, Stanwix P~L, Hodges J~S, Hong S, Taylor J~M, Cappellaro P, Jiang L,
  Dutt M~V, Togan E, Zibrov A~S, Yacoby A, Walsworth R~L and Lukin M~D 2008
  {\em Nature\/} {\bf 455} 644--647 ISSN 14764687

\bibitem{DeLange2011}
{De Lange} G, Rist{\`{e}} D, Dobrovitski V~V and Hanson R 2011 {\em Phys.Rev.
  Lett.\/} {\bf 106} ISSN 080802

\bibitem{Plenio2019}
Pedernales J~S, Morley G~W and Plenio M~B 2019 Motional dynamical decoupling
  for matter-wave interferometry (\textit{Preprint} \eprint{arXiv:1906.00835})

\bibitem{Scerri2020}
Scerri E, Gauger E~M and Bonato C 2020 {\em New J.Phys.\/} {\bf 3,22} 035002

\end{thebibliography}

\end{document}